\documentstyle[preprint,floats,tighten,prd,aps]{revtex}

\def \MSbar {\vbox{\hrule\kern 1pt\hbox{\rm MS}}}

\input epsf.tex
\def\DESepsf(#1 width #2){\epsfxsize=#2 \epsfbox{#1}}

\begin{document}

\title{QCD Calculations by Numerical Integration}
\author{ Davison E.
\ Soper}
\address{Institute of Theoretical Science, 
University of Oregon, Eugene, OR 97403}

\date{29 April 1998}
\maketitle

\begin{abstract}
Calculations of observables in Quantum Chromodynamics are typically
performed using a method that combines numerical integrations
over the momenta of final state particles with analytical integrations
over the momenta of virtual particles. I discuss a method for performing
all of the integrations numerically.
\end{abstract}



\narrowtext
This Letter concerns a method for performing perturbative calculations in
Quantum Chromodynamics (QCD) and other quantum field theories. Specifically, I
am concerned with cross sections and other QCD observables in which one
measures something about the hadronic final state. Here one cannot use the
special techniques that apply to inclusive quantities like the structure
functions in deeply inelastic lepton scattering. The general class of
calculations of interest in this Letter includes jet cross sections in
hadron-hadron and lepton-hadron scattering and in $e^+ e^- \to {\it 
hadrons}$. There have been many calculations of this kind  carried out at
next-to-leading order in perturbation theory, resulting in an impressive
confrontation between theory and experiment and in an accumulation of
evidence supporting QCD as the correct theory of the strong interactions
\cite{ESW}. These calculations are based on a method introduced by Ellis,
Ross, and Terrano \cite{ERT} in the context of $e^+ e^- \to {\it  hadrons}$.
Stated in the simplest terms, the Ellis-Ross-Terrano method is to do
some integrations over momenta $\vec p_i$ analytically, others numerically. I
shall argue that it is possible instead to do all of these integrations
numerically. Furthermore, I shall argue that performing all of the
integrations numerically has some advantages, principally in the flexibility
that it allows.

In this Letter, I address only the process $e^+ e^- \to {\it  hadrons}$.
Thus I do not address the issue of factorization that is associated with
initial state hadrons. I discuss three-jet-like infrared safe observables at
next-to-leading order, that is order $\alpha_s^2$. Examples of such
observables include the thrust distribution, the fraction of events that have
three jets, and the energy-energy correlation function.

Let us begin with a precise statement of the problem. The order $\alpha_s^2$
contribution to the observable being calculated has the form
\begin{eqnarray}
\sigma^{[2]} &=&
{1 \over 2!}
\int d\vec k_1 d\vec k_2\
{d \sigma^{[2]}_2 \over d\vec k_1 d\vec k_2}\
{\cal S}_2(\vec k_1,\vec k_2)
\nonumber\\
&&+
{1 \over 3!}
\int d\vec k_1 d\vec k_2 d\vec k_3\
{d \sigma^{[2]}_3 \over d\vec k_1 d\vec k_2 d\vec k_3}\
{\cal S}_3(\vec k_1,\vec k_2,\vec k_3)
\label{start}\\
&&
+
{1 \over 4!}
\int d\vec k_1 d\vec k_2 d\vec k_3 d\vec k_4\
{d \sigma^{[2]}_4 \over d\vec k_1 d\vec k_2 d\vec k_3 d\vec k_4}\
{\cal S}_4(\vec k_1,\vec k_2,\vec k_3,\vec k_4).
\nonumber
\end{eqnarray}
Here the $d\sigma^{[2]}_n$ are the  order $\alpha_s^2$ contributions to the
parton level cross section, calculated with zero quark masses. Each contains
momentum and energy conserving delta functions. The $d \sigma^{[2]}_n$ include
ultraviolet renormalization in the \MSbar\ scheme. The functions $\cal S$
describe the measurable quantity to be calculated. We wish to calculate a
``three-jet'' quantity.  That is, ${\cal S}_2 = 0$. The normalization is such
that ${\cal S}_n = 1$ for $n = 2,3,4$ would give the order $\alpha_s^2$
perturbative contribution the the total cross section. There are, of course,
infrared divergences associated with Eq.~(\ref{start}). For now, we may
simply suppose that an infrared cutoff has been supplied.

The measurement, as specified by the functions ${\cal S}_n$, is to be infrared
safe, as described in Ref.~\cite{KS}: the ${\cal S}_n$ are smooth functions of
the parton momenta and
\begin{equation}
{\cal S}_{n+1}(\vec k_1,\dots,\lambda \vec k_n,(1-\lambda)\vec k_n)
= 
{\cal S}_{n}(\vec k_1,\dots, \vec k_n)
\end{equation}
for $0\le \lambda <1$. That is, collinear splittings and soft particles do
not affect the measurement.

It is convenient to calculate a quantity that is dimensionless. Let the
functions ${\cal S}_n$ be dimensionless and eliminate the remaining
dimensionality in the problem by dividing by $\sigma_0$, the total $e^+ e^-$
cross section at the Born level. Let us also remove the factor of $(\alpha_s
/ \pi)^2$. Thus, we calculate
\begin{equation}
{\cal I} = {\sigma^{[2]} \over \sigma_0\ (\alpha_s/\pi)^2}.
\end{equation}

We note that ${\cal I}$ is a function of the c.m.\ energy
$\sqrt s$ and the $\overline{\rm MS}$ renormalization scale $\mu$. We will
choose $\mu$ to be proportional to $\sqrt s$: $\mu = A_{UV} \sqrt s$. Then
${\cal I}$ depends on $A_{UV}$. But, because it is dimensionless, it is
independent of $\sqrt s$. This allows us to write
\begin{equation}
{\cal I} = \int_0^\infty d \sqrt s\ h(\sqrt s)\ 
{\cal I}(A_{UV},\sqrt s),
\label{rtsintegral}
\end{equation}
where $h$ is any function with
\begin{equation}
\int_0^\infty d \sqrt s\ h(\sqrt s) = 1.
\end{equation}
The integration over $\sqrt s$ eliminates the energy conserving delta function
in $\cal I$. The physical meaning is that, by smearing in the energy
$\sqrt s$, we force the time variables in the two current operators that
create the hadronic state to be within $1/\sqrt s$ of each other. Thus we
have a truly short distance problem.

I now describe how one would calculate $\cal I$ using the Ellis-Ross-Terrano
method. Each partonic cross section in Eq.~(\ref{start}) can be expressed as
an amplitude times a complex conjugate amplitude. One must calculate the
amplitudes in $4 - 2 \epsilon$ dimensions. (In the case of the process
$e^+e^- \to {\it hadrons}$, this calculation was performed by  Ellis, Ross,
and Terrano \cite{ERT}.) For tree diagrams, the calculation is
straightforward. For loop diagrams, this involves an integration, which is
performed analytically. The integrals are divergent in four dimensions, so
one obtains divergent terms proportional to $1/\epsilon^2$ and $1/ \epsilon$
in addition to terms that are finite as $\epsilon \to 0$. Having the
amplitudes and complex conjugate amplitudes, one must now multiply by the
functions ${\cal S}_n$ and integrate over the final state parton momenta.
These integrations are too complicated to perform analytically, so one must
use numerical methods. Unfortunately, the integrals are divergent at
$\epsilon = 0$. Thus one must split the integrals into two parts. One part
can be divergent at $\epsilon = 0$ but must be simple enough to calculate
analytically. The other part can be complicated, but must be convergent at
$\epsilon = 0$. One calculates the simple, divergent part and cancels the
$1/\epsilon^2$ and $1/ \epsilon$ pole terms against the pole terms coming
from the virtual loop diagrams. This leaves the complicated, convergent
integration to be performed numerically.

This method is a little bit cumbersome, but it works and has been enormously
successful. However it has proven to be difficult to apply the method in the
case of two virtual loops. Even with one virtual loop, the method is not very
flexible. Any modification of the integrand requires one to recalculate the
amplitudes, and the modification must be simple enough that one can calculate
the amplitudes in closed form.

\begin{figure}
\centerline{\DESepsf(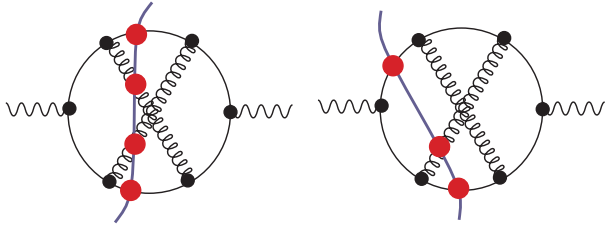 width 8 cm)}
\caption{Two cuts of one of the Feynman diagrams that contribute to $e^+e^-
\to {\it hadrons}$.}
\label{cutdiagrams}
\end{figure}

Let us, therefore, inquire whether there is any other way that one might
perform such a calculation. We note that the quantity $\cal I$ can be
expressed in terms of cut Feynman diagrams, as in Fig.~\ref{cutdiagrams}. The
part of the diagram to the left of the cut is a term in the amplitude. The
part to the right of the cut is a term in the complex conjugate amplitude.
The dots where the parton lines cross the cut are intended to represent the
function ${\cal S}_n(\vec k_1, \dots ,\vec k_n)$. Each diagram is a three loop
diagram, so we have integrations over loop momenta $\ell_1^\mu$, $\ell_2^\mu$
and $\ell_3^\mu$. 

We first perform the energy integrations. For the graphs in which four parton
lines cross the cut, there are four mass-shell delta functions $\delta
(k_J^2)$. These delta functions eliminate the three energy integrals over
$\ell_1^0$, $\ell_2^0$, and $\ell_3^0$ as well as the integral
(\ref{rtsintegral}) over $\sqrt s$. For the graphs in which three parton
lines cross the cut, we can eliminate the integration over $\sqrt s$ and two
of the $\ell_J^0$ integrals. One integral over the energy $E$ in the virtual
loop remains. The integrand contains a product of factors
\begin{equation}
{ i \over E - Q_J^0 - \omega_J + i\epsilon}\,
{ 1 \over E - Q_J^0 + \omega_J - i\epsilon},
\end{equation}
where $E - Q^0_J$ is the energy carried by the $J$th propagator around the
loop and $\omega_J$ is the absolute value of the momentum carried on that
propagator. We perform the integration by closing the integration contour in
the lower half $E$ plane. This leads to $n$ terms for a virtual $n$ point
subgraph. In the $J$th term, the propagator energy $E - Q^0_J$ is set equal to
the corresponding $\omega_J$ and there is a factor $1/(2\omega_J)$. Note that
the entire process of performing the energy integrals amounts to some
simple algebraic substitutions.

Let us denote the contribution to $\cal I$ from the cut $C$ of graph $G$ by
${\cal I}(G,C)$. This contribution has the form
\begin{equation}
{\cal I}(G,C) = \int d^3\vec \ell_1\,d^3\vec \ell_2\,d^3\vec \ell_3\
g(G,C;\ell),
\end{equation}
where $\ell$ denotes the set of loop momenta $\{\vec\ell_1, \vec\ell_2,
\vec\ell_3\}$. The functions $g$ have some singularities, called pinch
singularities, that cannot be avoided by deforming the integration contour
and some non-pinch singularities that can be avoided by a contour
deformation. A detailed analysis, which I sketch below, has been given by
Sterman \cite{sterman}. 

The pinch singularities occur when one parton branches into two partons
with collinear momenta or when one parton momentum goes to zero. These
singularities can lead to logarithmic divergences in the corresponding
integral. (As pointed out in \cite{sterman}, the $q^\mu q^\nu$ terms in
gluon self-energy subgraphs lead to {\it quadratic} divergences. I will
discuss these terms shortly.) If we were to calculate the total cross section
by using measurement functions ${\cal S}_n = 1$ in $g$, then the singularities
would cancel \cite{sterman} between the functions $g$ associated with the
various cuts $C$ of the same graph $G$. The underlying reason is unitarity.
Now in our case of measured shape variables, the values of ${\cal S}_n$
corresponding to different cuts $C$ are different. This would ruin the
cancellation, except that just at the collinear or soft points the functions
${\cal S}_n$ match.  Thus the singularities present in the individual
$g(G,C;\ell)$ cancel in the sum, $\sum_C g(G,C;\ell)$. To be precise, the
cancellation reduces the strength of the singularity from a strength
sufficient to give a logarithmically divergent integral to one that gives a
convergent integral.

The function $g$ also has singularities that can be avoided by deforming the
integration contours, the {\it non-pinch} singularities. These can be
characterized rather simply in the present case of a single virtual loop. Let
$\vec \ell$ be the a loop momentum that flows around this virtual
loop. We may choose $\vec \ell$ so that it is the momentum carried by one of
the loop propagators and so that a momentum $\vec Q$ flows out of the loop and
into the final state between the propagator in question and another propagator
further along the loop. This second propagator carries momentum $\vec
\ell - \vec Q$. A positive energy $Q^0$ also flows out of the loop and into
the final state. Since $(Q^0,\vec Q)$ is the four-momentum of a group of
final state particles, we have $Q^0\ge |\vec Q|$. Then there is a singularity
when $|\vec \ell| + |\vec \ell - \vec Q| = Q^0$. In the case $Q^0 = |\vec Q|$,
this is a collinear singularity, and will cancel between cuts. In the case
$Q^0> |\vec Q|$, the singular surface is an ellipsoid. The singularity does
not cancel, but the Feynman rules provide an $i\epsilon$ prescription that tells
us that we should deform the $\vec \ell$ integration contour into the complex 
$\vec \ell$ plane so as to avoid the singularity. Here deforming the contour
means replacing $\vec \ell$ by a complex vector $\vec \ell + i
\vec \kappa$. Then one simply chooses the imaginary part, $\vec \kappa$, of
the loop momentum as a function of the real part, $\vec \ell$, and supplies an
appropriate jacobian $\cal J$. Since the momentum $\vec \ell$ that flows
around the virtual loop in question is, in general, a linear combination of
the three loop momenta $\vec\ell_j$, one should write the general relation as
$\vec\ell_j \to \vec\ell_j + i\vec \kappa_j$, or, in a abbreviated notation,
$\ell \to \ell + i\kappa$. Then
\begin{equation}
{\cal I}(G,C) = \int d^3\vec \ell_1\,d^3\vec \ell_2\,d^3\vec \ell_3\
{\cal J}(G,C;\ell)\
g(G,C;\ell + i\kappa(G,C;\ell)).
\end{equation}
Note that the contour deformation is more than just a mathematical trick. The
analyticity that allows it is a consequence of the causality property of the
field theory.

Now, there is a simple argument that the pinch singularities cancel between
different cuts $C$ of the same graph $G$. There is another simple argument that
the non-pinch singularities are not a problem because one can deform the
integration contours so as to avoid them. If one is being careful in the
proof, one notes that the deformations required to escape the non-pinch
singularities are different for each cut $C$. (The deformations for
different cuts must be different if one wants the momenta of final state
particles to be real.) On the other hand, the cancellation of pinch
singularities requires that the contours for different cuts $C$ be the same as
one approaches the pinch singularities. That is, the set of vectors
$\{\vec\kappa_1,\vec\kappa_2,\vec \kappa_3\}$ that define the deformation
associated with a cut $C$ must to go to zero as $\{\vec \ell_1, \vec
\ell_2,\vec \ell_3\}$ approaches a pinch singular surface.

This sort of consideration of how to prove that $\cal I$ is finite
instructs us how to actually calculate $\cal I$. With the contours
appropriately chosen, the integral
\begin{equation}
{\cal I}(G) = \int d^3\vec \ell_1\,d^3\vec \ell_2\,d^3\vec \ell_3\
\sum_C\,
{\cal J}(G,C;\ell)\
g(G,C;\ell + i\kappa(G,C;\ell))
\label{master}
\end{equation}
is finite. One can simply compute it by Monte Carlo integration. Note the
significance of putting the summation over cuts inside the integral. When we
sum over cuts for a given point in the space of loop momenta, the soft and
collinear divergences cancel because the cancellation is built into the
Feynman rules. If we were to put the sum over cuts outside the integration, as
in the Ellis-Ross-Terrano method, then the individual integrals would be
divergent. The calculation would thereby be rendered more difficult.

Eq.~(\ref{master}) represents the main point of this Letter. It is perhaps
useful to point out that one must remove from the integration tiny regions
near the collinear and soft singular points, since otherwise roundoff errors
would spoil the cancellation of the individual contributions. I leave for
later, more specialized, papers the details of how one can choose a specific
contour deformation so that the theoretical cancellation is realized in
practice. I also skip a discussion of how one can choose the density of
integration points so as to perform the numerical integration by the Monte
Carlo method. There are, however, two points of principle that I discuss
below, since they are not analyzed in Ref.~\cite{sterman}: renormalization and
the special treatment required for self-energy subgraphs.

Renormalization is conventionally accomplished by performing loop integrations
in $4-2\epsilon$ space-time dimensions and subtracting the resulting pole term,
$c/\epsilon$. Clearly that is not appropriate in a numerical integration.
However, one can subtract instead an integral in 4 space-time
dimensions such that, in the region of large loop momenta, the integrand of the
subtraction term matches the integrand of the subgraph in question. The
integrand of the subtraction term can depend on a mass parameter $\mu$ in such
a way that the subtraction term has no infrared singularities. Then, one can
easily arrange that this {\it ad hoc}  subtraction has exactly the same effect
as \MSbar\ subtraction with scale parameter $\mu$.

Self-energy subgraphs require a special treatment. Consider a quark self-energy
subgraph $-i\Sigma$ with one adjoining virtual propagator and one adjoining
cut propagator. This combination really represents a field strength
renormalization for the quark field, and is interpreted as
\begin{equation}
{ 1 \over 2}
\left[{ \rlap{/}q \over q^2}\,\Sigma(q)\,\rlap{/}q\right]_{q^2 = 0}\, 2\pi 
\delta(q^2).
\end{equation}
In order to take the $q^2 \to 0 $ limit here while at the same time
maintaining the cancellation of collinear divergences, we write
\begin{equation}
{ \rlap{/}q\,\Sigma(q)\,\rlap{/}q \over q^2}
= - { g^2 \over (2\pi)^3}\,C_F
\int d^3\vec\ell\
{ q^0 |\vec k_+|\gamma^0 - (|\vec k_+|+|\vec k_-|)\vec k_+\cdot \vec \gamma
\over |\vec k_+| |\vec k_-|\left[(|\vec k_+|+|\vec k_-|)^2 - (q^0)^2\right]},
\label{dispersion}
\end{equation}
where $\vec k_\pm ={ 1 \over 2}\vec q \pm \vec \ell$. This expression is
obtained by writing the left-hand side as a dispersive integral with the cut
self-energy graph appearing as the discontinuity. When the virtual
self-energy is written in this form, the $q^2 \to 0$ limit is smooth and, in
addition, the cancellation between real and virtual graphs in the collinear
limit $\vec \ell \propto \vec q$  is manifest. It should be noted that the
integral in Eq.~(\ref{dispersion}) is ultraviolet divergent and requires
renormalization, which can be performed with an {\it ad hoc}
subtraction as described above.

One may expect that the representation of virtual propagator corrections in
terms of the cut propagator will prove to be convenient in future
modifications of the method described in this Letter. One may want to make
modifications to the gluon propagator, in particular, in order to implement a
running strong coupling and to insert models for the long distance propagation
of gluons. In addition, one may want to modify propagators so
that partons can enter the final state with $q^2 >0$ so as to mesh an
order $\alpha_s^2$ perturbative calculation with a parton shower Monte Carlo
program.

Consider, finally, the one loop gluon self-energy subgraph, $\pi^{\mu\nu}(q)$.
The term in $\pi^{\mu\nu}(q)$ proportional to $q^\mu q^\nu$ contributes
quadratic infrared divergences \cite{sterman}. This problem can be
addressed by replacing  $\pi^{\mu\nu}$ by $P^\mu_\alpha\pi^{\alpha\beta}
P_\beta^\nu$, where
$
P^\mu_\alpha = g^\mu_\alpha - { q^\mu \tilde q_\alpha / \tilde q^2},
$
with $\tilde q = (0,\vec q)$. The terms added to $\pi^{\mu\nu}$ are
proportional to either $q^\mu$ or $q^\nu$ and thus vanish when one sums over
different ways of inserting the dressed gluon propagator into the remaining
subgraph. Since $P^\mu_\alpha q^\alpha = 0$, the problematic $q^\mu
q^\nu$ term is eliminated. Effectively, this is a change of gauge for dressed
gluon propagators from Feynman gauge to Coulomb gauge.
 
\begin{figure}
\centerline{\DESepsf(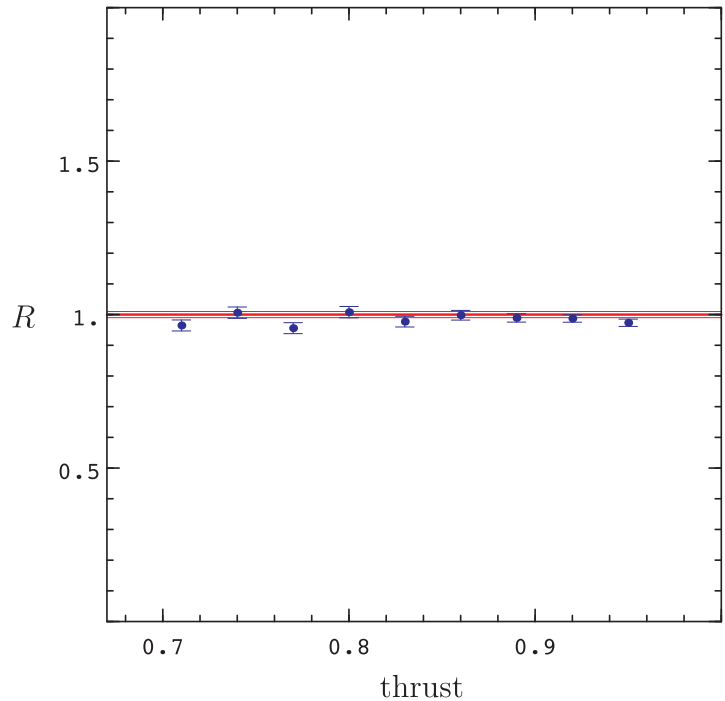 width 10 cm)}
\caption{Ratio $R$ of the order $\alpha_s^2$ contribution to the thrust
distribution as calculated here to the same quantity as calculated by Kunszt
and Nason, Ref.~\protect\cite{KN}. The horizontal lines represent the
expected result, 1, with an error estimate. The error bars on the computed
points represent statistical errors. There is some correlation expected
between neighboring points.}
\label{answer}
\end{figure}

We have seen that a completely numerical integration of the cut Feynman
diagrams for a physical quantity can, in principle, produce the numerical
value of the quantity. Furthermore, there may be advantages in simplicity and
flexibility associated with this approach. The question naturally arises,
can such a calculation work, in a practical sense? In order to find
out, I have constructed a demonstration computer program along the lines
outlined above \cite{beowulf}. I have used the program to calculate
$d\sigma^{[2]}/ d T$, the order $\alpha_s^2$ contribution to the thrust
distribution. More precisely, I have calculated the ratio $R(T)$ of
$d\sigma^{[2]}/ d T$ to $[d\sigma^{[2]}/ d T]_{\rm KN}$, where
$[d\sigma^{[2]}/ d T]_{\rm KN}$  is a fit to the tabulated results for
$d\sigma^{[2]}/ d T$ as given by Kunszt and Nason \cite{KN}. In the range
$0.71<T<0.95$, the function $[d\sigma^{[2]}/ d T]_{\rm KN}$ varies by about a
factor of 30. The ratio $R(T)$ should be 1. The results are reported in
Fig.~\ref{answer}.  We can conclude that the automatic cancellations between
different cuts of the same diagram are indeed realized and that completely
numerical integration for QCD observables beyond the leading order is a
practical possibility.

{\it Outlook.} Substantial effort will be required to test the computer code
used for Fig.~\ref{answer} in order to detect and remove any bugs that it may
contain and then to document the code and the algorithms used. This work will
be reported in future papers. With this code in hand, or with improved code
from other authors, one can attack more difficult problems than discussed
here. It remains to be seen for what problems the completely numerical method
will prove to be more powerful than the analytical/numerical method that has
served us so well up to now.

I thank U.~Amaldi for initial encouragement to somehow do better with QCD
calculations, L.~Surguladze for advice on Feynman numerators,
Z.~Kunszt and P.~Nason for help with the comparison to the results
\cite{KN}, and T.~Sjostrand for advice about future applications. This work
was supported by the U.\ S.\ Department of Energy.

\end{document}